# Shells of crystal field symmetries evidenced in oxide nano-crystals

**B Masenelli[1], G Ledoux[2], D Amans[2], C Dujardin[2] and P Mélinon[3]**

[1]Université de Lyon, F-69000 Lyon, France and INL, CNRS, UMR 5270, INSA Lyon, F-69621 Villeurbanne, France

[2]Université de Lyon, Lyon, F-69000, France; Université Lyon 1, Villeurbanne, F-69622, France; CNRS, UMR5620, LPCML, Villeurbanne F-69622, France

[3]Université de Lyon, Lyon, F-69000, France; Université Lyon 1, Villeurbanne, F-69622, France; CNRS, UMR5586, LPMCN, Villeurbanne F-69622, France

E-mail : bruno.masenelli@insa-lyon.fr

**Abstract.** By the use of a point charge model based on the Judd-Ofelt transition theory, the luminescence from $Eu^{3+}$ ions embedded in $Gd_2O_3$ clusters is calculated and compared to the experimental data. The main result of the numerical study is that without invoking any other mechanisms such as crystal disorder, the pure geometrical argument of the symmetry breaking induced by the particle surface has influence on the energy level splitting. The modifications are also predicted to be observable in realistic conditions where unavoidable size dispersion has to be taken into account. The emission spectrum results from the contribution of three distinct regions, a cluster core, a cluster shell and a very surface, the latter being almost completely quenched in realistic conditions. Eventually, by detailing the spectra of the ions embedded at different positions in the cluster we get an estimate of about 0.5 nm for the extent of the crystal field induced Stark effect. Due to the similarity between $Y_2O_3$ and $Gd_2O_3$, these results apply also to $Eu^{3+}$ doped $Y_2O_3$ nanoparticles.

## 1. Introduction
Complex oxides (rare earth sesquioxides, ternary or quaternary compounds [1,2,3,4,5]) doped with luminescent rare earth ions, such as $Eu^{3+}$, are attractive candidates for the design of nanometric light sources. They are often highly stable under high power excitation and allow choosing the emission wavelength through the choice of the doping element. Their use as nanoparticles is justified by the need of miniaturization of photonic devices as well as of biological systems in which they serve as labels. However, as the size is reduced down to a few nanometers in diameter (typically 5 or 6 nm), the spectrum of the doping rare earth element, ordinarily exhibiting very sharp peaks, tends to show inhomogeneously broadened peaks [1,6,7,8,9], a phenomenon not completely understood yet. The emission of trivalent rare earth ions in complex oxide matrices often results from transitions between f-f states. Due to the shielding effect and therefore the low electron-phonon coupling of these states, these transitions are composed of sharp lines. Accordingly, any change of the crystal field induces small line shifts and branching ratio changes easy to detect. Consequently, the inhomogeneous broadening of these transitions has been tentatively explained by the appearance of crystal disorder in the matrices network as the materials size is reduced [1]. However, *ab initio* calculations combined to transmission electronic microscope observations [10] have shown that the high ionicity of these compounds forbids the crystal disorder as long as the compound is strictly stoichiometric. Nevertheless, some disorder may occur in nanogranular films, where interfaces between particles act as grain boundaries and where the Gibbs pressure induced by the surface for very small particles can create an inhomogeneity of the lattice among the particle ensemble [10]. Besides, as the size is reduced, the relative amount of surface atoms increases. Since they are sites favoured for adsorption (and oxides do adsorb water and $O_2$ among others [11,12]), the luminescence of ions at the very surface may be strongly affected, causing the broadening evoked above.

Both explanations are valid. In particular, Figure 1 exemplifies the importance of surface contamination on the emission spectrum of $Eu^{3+}$ ions embedded in $Gd_2O_3$ nanoparticles. Figure 1 shows the photoluminescence spectra [see appendix] of the same sample of clusters synthesized by the LECBD technique [11,13] (low energy cluster beam deposition) which has been progressively annealed in order to increase the mean diameter of the particles (details of this procedure are presented in reference [14]). During annealing, the quantity of matter thus remains constant. The bottom curve of the as-deposited clusters presents large peaks due to the mixed composition of the sample: cubic and monoclinic phases [15]. The same behavior is observed on $Gd_2O_3$ nanoparticles whatever the synthesis methods, including soft chemical routes [16], low-energy cluster beam deposition [14-15], laser pyrolysis [17], combustion synthesis [9] and pulsed laser ablation in liquids [18]. This spectroscopic observation is thus a robust size effect that does not depend much on the surface of particles. At the beginning of the annealing (300°C for 1 hour, second curve from the bottom), despite the fact that the clusters retain their mean size (the temperature is not high enough to promote sintering), the overall emission intensity increases by one order of magnitude. In this step, neither crystal modification nor cluster coalescence occurs. Only desorption of contamination species takes place. This illustrates the importance of the surface contamination on the luminescence quenching. Note, however, that the spectrum is still broadened. In particular, a broad peak between 615 and 630 nm appears when the surface is cleaned. When the size of the particles increases, the peaks in the spectrum get sharper and the total intensity (total efficiency) concomitantly increases by another order of magnitude. Eventually, the spectrum of the 10 nm clusters is almost tantamount to the bulk crystal one.

The broadening of the uncontaminated clusters spectrum, also observed for particles kept in ultra high vacuum [15] goes to show that there must be another reason at stake. In order to investigate on this phenomenon, we have numerically simulated the effect of size reduction on the Stark effect induced by the matrix field and compared the results to experimental results.

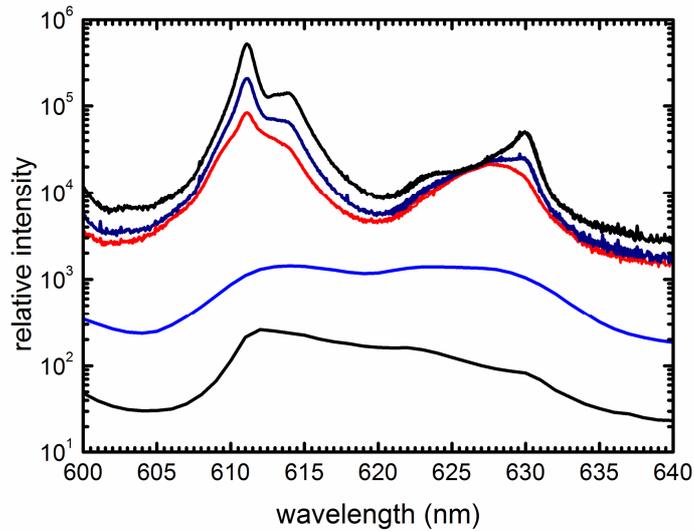

**Figure 1 : experimental $^5D_0 \rightarrow {}^7F_2$ emission of $Eu^{3+}$ ions (10%) embedded in $Gd_2O_3$ clusters of different sizes (photoluminescence experiment at room temperature; excitation at 260 nm). The mean size of the clusters has been increased by annealing (see ref. [14] for details). The quantity of matter probed by the excitation beam is identical for all spectra, allowing direct comparison between the relative intensities of the spectra. From bottom to top : as-deposited clusters with a mean diameter of 3.2 nm; as-deposited clusters heated at 300°C (no coalescence, removal of surface contamination); 4 nm; 5 nm; 9 nm.**

## 2. Code description

There are many models used for the calculation of the energy levels of a rare earth ion in a matrix: the electrostatic model [19], the angular overlap model [20], the molecular orbital model [21] and *ab initio* methods [22]. In the present work, a full calculation of the crystal field within the point charge approximation is performed. The ion energy levels as well as the transition probabilities are calculated within the framework of a code developed by Monteil and co-workers [23,24]. Briefly, the spectral characteristics $^5D_0 \rightarrow {}^7F_j$ (j = 0-7) of the $Eu^{3+}$ ion in a $Gd_2O_3$ matrix are calculated using a complete treatment of the crystal field assuming a point charge distribution and the radiative transition intensities are calculated according to the Judd-Ofelt theory [25,26]. The Stark effect on the 4f levels of the dopant is treated in the framework of the perturbations theory. The point charge assumption neglects both the spatial spreading of the charge density of the surrounding ions and the overlap between the optically active 4f wave functions and those of the ions. This method, developed by Leavitt, Morrison *et al.* [27,28], has already proven its efficiency in many studies [23,24,29,30] and despite its limitations is well suited for treating highly ionic systems such as $Gd_2O_3$ and other rare earth oxides. On a practical point of view, three parameters need to be set before simulations. First, the optical index of the medium surrounding the doping $Eu^{3+}$ ions has been set to 1.5. Its theoretical knowledge requires the description of the material on a length scale of the order of the wavelength [31] of the doping element emission (in our case, approximately 611 nm). Its experimental value can be obtained by ellipsometry. But in nanogranular samples with an important porosity, the analysis is often awkward. Instead, we have checked that changing the value of the optical index between 1 and 2 had negligible effect (cf. supporting data). Therefore we have kept the value of 1.5 used for gadolinium oxide glasses, which is intermediate between 1 (vacuum) and 1.98 for bulk $Gd_2O_3$. Second, we have set the actual valence of oxygen and gadolinium (europium) ions to 1.5 and 2.25 respectively. The valence values have a direct effect on the wavelength of the $^5D_0 \rightarrow {}^7F_j$ transitions. Their accurate determination is therefore straightforward through the comparison of the predictions to experimental data for $Eu^{3+}$ doped bulk $Gd_2O_3$ (cf. supporting data). Furthermore, the deduced values are in good agreement with values calculated *ab initio* by Xu *et al.* [32] for $Y_2O_3$ (a material identical to $Gd_2O_3$). Eventually, we set the full width at half maximum of each transition to 8 cm$^{-1}$ based on the experimental resolution of our setup. Note that all the ions constituting a nanoparticle are included in the calculation of the crystal field at a given location among the nanoparticle. Spherical nanoparticles have been considered for the sake of simplicity, even though we have previously demonstrated that naked nanoparticles of rare earth sesquioxide tend to adopt rhombic dodecahedron shape [15]. We have checked that the choice of geometry has no significant influence. Besides, the doping level of the clusters is low enough (10% of Gd substituted with Eu, below the percolation threshold) to consider the experimental dopant distribution as homogeneous. This has been the case in our simulations too. The positions of the ions within the cluster have been kept to the nominal crystallographic value. This is justified by calculations that have revealed that a highly ionic nanoparticle does not significantly modify its crystal network. Only ions at the corners or edges of a facetted particle are affected by a structural relaxation (contraction motion) [10].

## 3. Simulation of the emission of $Eu^{3+}$ ions embedded in $Gd_2O_3$ clusters

### 3.1. Emission from a unique particle: range of Stark effect

To get some insight into the range of the Stark effect induced by the crystal field in a $Gd_2O_3$ nanoparticle, we have plotted in Figure 2 for a 4 nm particle the emissions of $Eu^{3+}$ ions located at different positions within the particle. Three distinct contributions are to be distinguished. The first one comes from the ions within the particle core, defined as the volume which radius equals the total radius minus 0.5 nm. $Eu^{3+}$ ions within this core exhibit a spectrum close to the $Gd_2O_3$ bulk one (cf. supporting data) and are thus not sensitive to the presence of the surface. A second contribution comes from ions contained in the remaining 0.5 nm thick shell minus the very surface (ions with half neighbours). In this region, ions have an inhomogeneously broadened spectrum. We can still distinguish the $^5D_0 \rightarrow {}^7F_2$ transition around 611 nm but other shifted contributions of the same transition appear. Eventually, the contribution from the ions at the very surface (0.2 nm thick shell, ions with half neighbours) appears to be most affected. In the latter case, the spectrum is significantly modified and looks like the one from rare earth doped glasses [23,24] or low symmetry phases (such as the monoclinic phase).

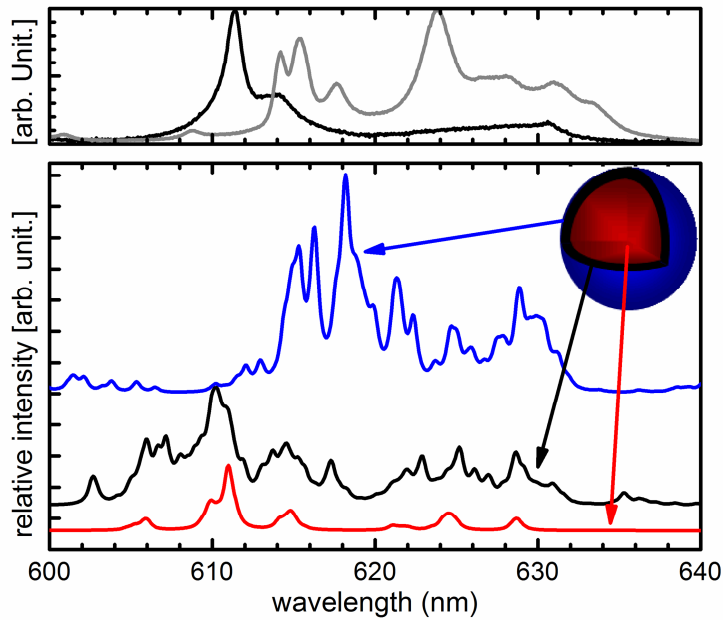

**Figure 2:** Top panel shows the experimental luminescence spectra (resolution 0.5 nm) of europium doped $Gd_2O_3$ in the cubic bulk phase (black line) and in the monoclinic bulk phase (gray line) [18]. Bottom panel shows calculated $^5D_0 \rightarrow {}^7F_j$ (j=1, 2) emissions of $Eu^{3+}$ ions embedded at different positions in a 4 nm $Gd_2O_3$ cluster. From top to bottom: ions at the very surface (with half of the nearest neighbours, blue curve), ions in a 0.5 nm thick shell (minus the surface contribution, black curve), ions in the cluster core (the cluster volume minus the 0.5 nm shell and the very surface, red curve). The spectra intensities can be compared and are **proportional to the number of emitting ions and thus to the volume considered. Inset: schematic view of the three regions.**

This result (the distinction of three zones) is valid whatever the diameter of the particle. Since these spectra result from the Stark effect induced by the crystal field at the $Eu^{3+}$ ions location which behave as crystal probes, our result implies that the crystal field is not perturbed inside the core of a nanoparticle but at its surface. This is not straightforward since the crystal field in ionic compounds is the sum of all the Coulomb fields of the ions surrounding a given site and thus may not converge quickly to a precise value as a function of the truncation of the sum. To estimate more precisely the range over which the crystal field induced Stark effect can be defined, we have plotted in Figure 3 the intensity ratio of the emissions at 611 nm and at 618 nm of $Eu^{3+}$ ions embedded in different shells of a 4 nm $Gd_2O_3$ nanoparticle. Three shells have been considered: the first one is the surface shell located from 0 to 0.2 nm from the surface, the second one between 0.2 and 0.5 nm and the last one between 0.5 and 1 nm from the surface. The core of the nanoparticle here is a sphere with a 1nm radius. The emission at 611 nm is representative of a perfect cubic bulk-like material whereas that at 618 nm is related to low symmetry sites. It is clearly seen that for ions embedded in shells distant from less than 0.5 nm from the surface the ratio is rather low revealing the influence of the symmetry breaking induced by the surface on the crystal field. For the shell located at more than 0.5 nm from the surface, the ratio is identical to the core one. Consequently, we can state that the Coulomb crystal field is defined over sizes as small as a 0.5 nm namely half a unit cell or equivalently two interatomic distances.

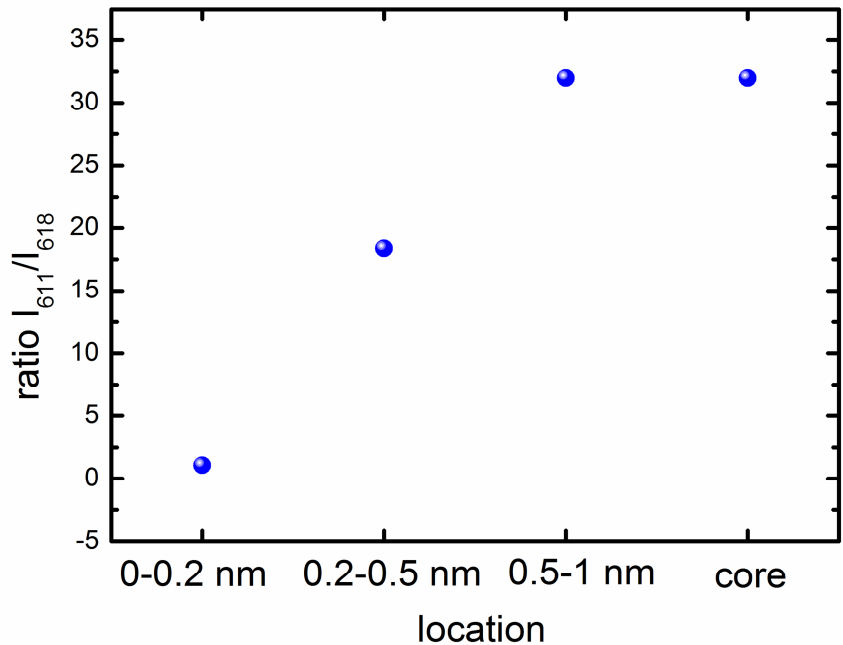

**Figure 3 : calculated intensity ratio of the emissions at 611 nm and at 618 nm of $Eu^{3+}$ ions embedded in different shells of a 4nm $Gd_2O_3$ nanoparticle. The location of the shells is defined from the surface. The core represents the 1nm radius sphere left inside the original particle.**

In order to compare the numerical predictions to experiments, we have depicted in Figure 4 the calculated integrated spectra of $Eu^{3+}$ ions in the same spherical $Gd_2O_3$ particle of 4 nm in diameter as a function of their location. More precisely, the spectra are obtained by summing the contribution of the $Eu^{3+}$ ions embedded in a core of variable thickness. At first glance, comparing the spectra from Figure 4 to those of Figure 1 we can state that the emission from the surface which dominates the spectrum of the entire particle is not observed for particles studied at air. The only spectrum from Figure 1 that looks like the prominent one in Figure 4 (with all the ion contributions) is the one obtained after a mild annealing meant to remove the surface contamination. We can thus conclude that most of the ions at the surface are optically inactive because of reaction with air moisture. Consequently the only major contributions that can be observed in standard conditions are those from the core (unperturbed) and from the 0.5 nm shell.

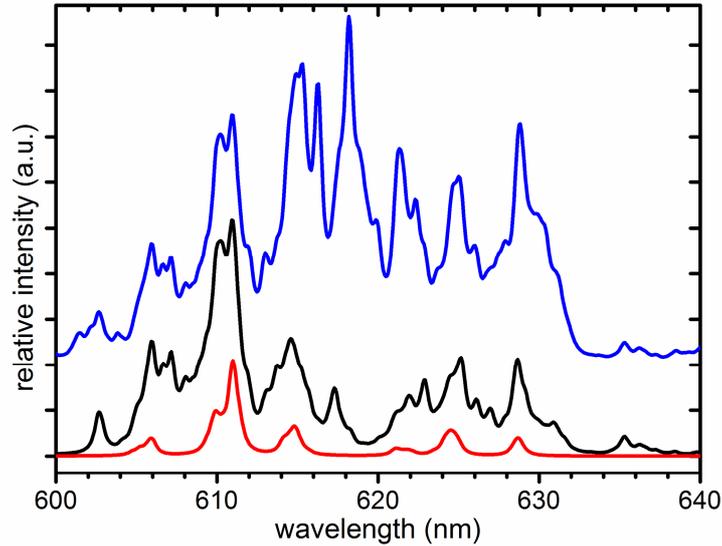

**Figure 4 : calculated $^5D_0 \rightarrow {}^7F_j$ emissions of $Eu^{3+}$ ions embedded in a 4 nm $Gd_2O_3$ cluster. From top to bottom: emission from all the ions (blue curve), from all the ions except those of the very surface (0.2 nm thick shell, black curve), from the ions in the cluster core (the cluster volume minus a 0.5 nm shell, red curve).**

*3.2. Comparison with experiments*

To confirm this statement, we have sought to calculate the emission spectra of samples as close as possible to the experimental samples. Any kind of samples presents size dispersion and we may thus wonder whether the dispersion would average out any possible effect of the size reduction on the Stark effect.

Figure 5 (top) exposes the experimental spectra of ensembles of $Gd_2O_3$ nanoparticles doped with 10% $Eu^{3+}$ ions with a mean diameter of 4 nm or 5 nm along with the spectrum of bulk $Eu^{3+}$ doped $Gd_2O_3$. Other results from the literature are presented in supporting data to illustrate the size-related emission broadening. The initial particles that form the nanostructured film have a size distribution that follows a log-normal law typical of the accretion process of the LECBD. After annealing to increase the mean size (cf. Figure 1) the size distribution is probably affected but cannot be estimated [15]. Since the annealing conditions are mild (at 450°C approximately, see [14]), we do not expect a drastic change. However the ripening of the clusters is a stochastic phenomenon. Thus we further assume that the distribution function turns into a normal law, typical of such processes. Figure 5 (bottom) also presents the calculated spectra for numerical samples close to the experimental ones assuming a normal law for the size dispersion. Since we only know the mean value of the distribution, we have set the standard deviation to the initial one, namely 2 nm. The aim of this part of the study is not to get a fit to the experimental curves but rather to get a qualitative insight into the modification of the crystal lattice induced Stark effect resulting from the finite size of the particles. To check to importance of the choice of the size distribution function, we have also calculated the spectra assuming a log-normal law. The results, presented as supporting data, do not show significant differences. Despite the absence of any fitting parameter, it can be observed that the calculated spectra are in relatively fair agreement with the experimental ones. In particular, the trend to broadening of the emission at 611 nm is reproduced as well as that of the emission between 620 nm and 630 nm. The simple fact of having a finite size can impact the emission of rare earth ions in rare earth sesquioxide nanoparticles in a significant way even when size dispersion is present.

However, the simulation is not quantitatively perfect. The calculations predict an emission at 606 nm which is not observed experimentally. Also the calculated peaks are sharper than the experimental ones. In particular, the shape of the contribution between 620 nm and 630 nm is not clearly reproduced but this is an intrinsic limitation of the code. This contribution is already not well reproduced for the bulk material (cf. supporting data). This could arise from the point charge model used in the calculations. A more sophisticated description of the charge distribution might improve the accuracy of the simulation (as *ab initio* calculations). Besides, the observed discrepancy may indicate the presence of a slight crystal disorder or an inhomogeneous surface passivation by air moisture. Indeed, if all the emitting ions at the surface are not passivated, the spectra of the particle ensemble will look much like the one of the unpassivated 4nm particle presented on Figure 4 (top curve). This is obviously not the case meaning that most of the surface ions are quenched. Nevertheless we cannot exclude that some among them are not completely quenched. They will thus contribute to the total spectrum with additional peaks leading to a broadening of the already present peaks. One last hypothesis is the fact that the surrounding particles may affect the emission from one particle by their Coulomb potential. This has not been calculated but may be of some importance especially for the ions in the surface shell. However, this is a difficult task since the result will depend on the choice of the positions and orientations of the considered particles and an average over these parameters would be required.

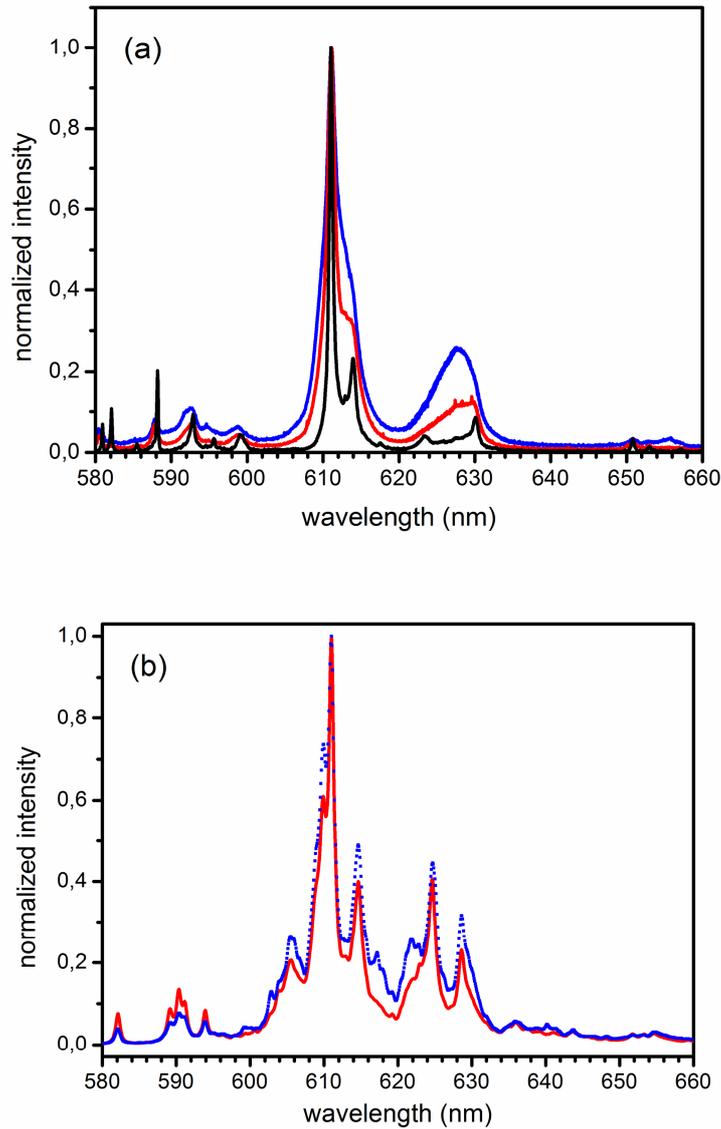

**Figure 5** (a): experimental $^5D_0 \rightarrow {}^7F_2$ emission of $Eu^{3+}$ ions embedded in $Gd_2O_3$ clusters with a mean diameter of 4 nm (blue curve), 5 nm (red curve) and in bulk from top to bottom respectively (black curve) (excitation wavelength of 260 nm); (b): calculated $^5D_0 \rightarrow {}^7F_2$ emission of $Eu^{3+}$ ions embedded in $Gd_2O_3$ clusters following a normal law of size distribution with a mean diameter of 4 nm (dotted line) and 5 nm (solid line). The spectra are normalized at their main peak. The broadening due to the size dispersion is taken into account in the calculations.

### 4. Conclusion

In conclusion, we have simulated by a point charge model based on the Judd-Ofelt transition theory the luminescence from $Eu^{3+}$ ions embedded in $Gd_2O_3$ clusters and compared the resulting spectra to previously obtained experimental spectra from clusters assembled films grown by the LECBD technique. The main result of the study is that in the spectrum of each cluster three contributions can be distinguished: a first one arising from the cluster core, a second one from a 0.5 nm thick shell and a third one from the ions at the very surface. The last contribution is often not observed experimentally because of its quenching by surface contamination. As the cluster size is reduced, the proportion of the shell according to the core increases so that at small sizes (less than 5 nm in diameter) the emission at 611 nm is inhomogeneously broadened. Other contributions at larger wavelengths appear as well, as a consequence of the lowering of the local symmetry.

Furthermore, the thickness of the shell in which the $Eu^{3+}$ emission is perturbed gives an estimate of the extent of the lattice crystal field induced Stark effect. This extent is of the same order as the shell thickness, namely 0.5 nm, which is about half the size of the lattice parameter. In other words, the crystal field is affected by the presence of the particle surface over two interatomic distances. Our calculations show that without invoking any other mechanisms such as crystal disorder, the pure geometrical argument of the symmetry breaking induced by the surface can lead to observable modifications of the spectra even in realistic conditions where a dispersion of the size particle has to be accounted for. This mechanism obviously does not exclude the other ones but rather is superimposed to the others (crystal disorder…). Besides, we emphasize that because of the similarity between $Y_2O_3$ and $Gd_2O_3$, these results apply also to $Eu^{3+}$ doped $Y_2O_3$ nanoparticles. To a certain extent, the perturbation of the crystal field and the resulting Stark effect due to the presence of the surface can be generalized to the emission of 4f transitions of rare earth ions in highly ionic nanoparticles.

**Acknowledgments**

The authors are indebted to Dr. D. Nicolas and Dr. B. Mercier for their support at the initial stage of the present work. This work has been supported by the French National Agency (ANR) in the frame of its program in Nanosciences and Nanotechnologies (NAPHO project n°ANR-08-NANO-054).

**Appendix**

Photoluminescence spectra have been performed on a homemade spectrofluorimeter. The samples were placed on a Linkham THMS600 heating plate that can heat up to 600°C. The excitation source is a 150W Xe lamp filtered through a Jobin Yvon Gemini double monochromator with resolution fixed at 2nm. The emitting light from the sample is collected by an optical fiber and fed to a Jobin Yvon TRIAX320 monochromator coupled with a Peltier cooled CCD. The system resolution was fixed at 0.25nm by using a grating of 1200gr/mm and slits of 100µm. Samples were heated at a given annealing temperature and then cooled down at room temperature before taking emission spectra.